\documentclass[12pt]{article}

\usepackage{mathptmx}
\usepackage{amsmath,amsthm,amssymb}
\usepackage{graphicx}
\usepackage{float}
\usepackage{lmodern,microtype}
\usepackage[margin=1in]{geometry}
\usepackage{kpfonts}	

\usepackage[round]{natbib}
\newcommand{\cmmnt}[1]{}
\usepackage{setspace}
\usepackage{tabularx}
\usepackage{subcaption}
\usepackage{nicefrac}
\usepackage{etoolbox}
\usepackage{xcolor}
\usepackage{tikz}
\usepackage{hyperref}

\usepackage{url}

\def\co{{\rm co}\hskip 1pt}

\def\proj{{\rm proj}}
\def\proj0{{\rm proj}_{\co w_0}}

\DeclareMathOperator{\diag}{diag}

\newtheorem{theorem}{Theorem}
{\end{theorem}\vskip.2cm}
\newtheorem{lemma}{Lemma}
{\end{lemma}\vskip.2cm}
\newtheorem{corolla}{Corollary}
{\end{corolla}\vskip.2cm}
\newtheorem{defini}{Definition}
{\end{defini}\vskip.2cm}
\newtheorem{proposi}{Proposition}
{\end{proposi}\vskip.1cm}

{\end{prop}\vskip.1cm}
\newtheorem{cla}{Claim}
{\end{cla}\vskip.2cm}
\newtheorem{assump}{Assumption}
{\end{assump}\vskip.1cm}
\newtheorem{hypoth}{Assumption}

{\end{hypoth}\vskip.1cm}
\newtheorem{demo}{Proof:}

{\cqfd \end{demo}}
\newtheorem{remark}{Remark}
{\end{remark}\vskip.3cm}

\usepackage[english]{isodate}
\graphicspath{ {./Image/} }
\usepackage{lineno}

\title{\Large \textbf{How Vulnerable is India's Economy to Foreign Sanctions?}}

\author{Vipin P. Veetil\thanks{Economics Area, Indian Institute of Management Kozhikode, Kerala 673 570, India.\newline \hspace*{0.6cm}I thank Ashwin Bhattathiripad for able research assistance.}}

\date{\small\printdayoff\today}

\begin{document}
\maketitle
\begin{abstract}
\setstretch{1.3}
This paper develops a simple model of the world supply chain to estimate the effects of sanctions
that restrict the flow of inputs from one country to another. Such restrictions operate through
changes in the weights of the global production network: the sanctioning country ceases supplying
certain inputs to the target country and reallocates its production to other destinations. Using the
OECD Inter-Country Input--Output tables, we calibrate the model to assess the vulnerability of the
Indian economy. We consider two classes of counterfactuals: restrictions on a single sector of a
foreign country supplying India, and restrictions on all sectors of a foreign country supplying
India. We then rank foreign countries and foreign country-sectors by the risk that their supply
restrictions pose to economic activity in India. Our results show that India's greatest
country-level vulnerability is to China, followed by the United Arab Emirates, the United States, Saudi Arabi and Russia, with the vulnerability to China being twice as much that to the UAE.
\end{abstract}


\newpage
\setstretch{1.4}
\section{Introduction}
One measure of interdependence between national economies is the ratio of world exports to world GDP. In 1830, this ratio was about \(6\%\); today it is about \(30\%\). Another measure of international production interdependence is the foreign value-added content of exports, or the share of exports that embodies imported intermediate inputs. By this metric too, cross-country production linkages increased markedly after 1990. Yet such measures of dependence likely understate the vulnerability of national economies to disruptions in international input flows. In several strategically important inputs, global supply is heavily concentrated in a small number of countries: China is the dominant exporter of natural graphite, Chile accounts for roughly two-thirds of lithium carbonate exports, and the Democratic Republic of the Congo remains overwhelmingly important in cobalt\footnote{For data on graphite, lithium and other critical minerals see \citet{WITSGraphite2024,WITSLithium2024,IEACriticalMinerals2025}. For exports and trade related statistics, see \citep{FedericoTenaJunguito2019,WorldBankExportsGDP,WorldBankWDR2020}} . In such a skewed setting, we must be able to study the consequences of restrictions imposed by a small number of countries---or even a single country---on the export of specific inputs.

\smallskip
The aforenoted vulnerabilities are not hypothetical. While the global economy is an
integrated whole, the political order that underwrites it remains a mosaic of 193 fragments of
varying size, most with armed forces of their own. Nor are relations among these
autonomous and semi-autonomous political entities uniformly peaceful. In 2024, there were \(61\)
active state-based armed conflicts across \(36\) countries, the highest number recorded since 1946,
and \(11\) of these reached the threshold of war. Economic coercion is likewise pervasive. Since
1950, there have been well over a thousand episodes of sanctions; this, without counting threats that were never imposed.\footnote{The conflict
figures are from the Uppsala Conflict Data Program. For sanctions, the most recent release of the
Global Sanctions Database records \(1{,}547\) sanctions cases over 1950--2023. For sanction threats,
the Threat and Imposition of Economic Sanctions (TIES) dataset records \(1{,}412\) sanctions cases
over 1945--2005, of which \(845\) involved actual sanctions and \(567\) involved threats without
imposition.} What we have, then, is an interdependent world economy built on
a `politically dynamic landscape', if I may be permitted that euphemism.

\smallskip
It is worth noting that the ultimate consequences of sanctions are not easy to gauge. Indeed, at the
moment sanctions are imposed, neither the sanctioning country nor the target country may know their
full effects; these reveal themselves only over time. One reason is the pliability of human
behavior: individuals, firms, and governments adjust to new circumstances in varied and often
creative ways, and such adjustments contain an element of irreducible unpredictability
\citep{buchanan1991market}. But there is also a second reason, one that is more mechanical. The
sensitivity of an economy to restrictions on the inflow of particular inputs depends not only on the
extent to which the domestic sectors using those inputs can substitute toward alternatives, but also
on how important those sectors are as suppliers of inputs to other sectors, and so on along the
chain of intersectoral dependence.  Furthermore, insofar as sanctions do not reduce the productive capacity of the sanctioning country,
they do not simply destroy resources; they redirect them from one country to another. That
redirection alters the international flow of goods and may increase the downstream availability of
certain inputs to countries not under sanction, and perhaps even, indirectly, to the country under
sanction itself. For instance, a sanction on the supply of coal to one country may increase that
country's access to steel if one of its steel suppliers gains greater access to coal as a result of
the very sanction. More generally, what matters is not merely the change in the flow of one input
from one nation to another, but the manner in which that disturbance reshapes the wider structure of
production.

\smallskip
The economic costs of sanctions, therefore, cannot be estimated without a coherent model of the flow of intermediate inputs through an integrated world production network. Which is precisely what we develop in this paper.  Our framework builds on a variant of the standard
buyer--seller production-network model, but works with the supplier-side dual of the usual
Cobb--Douglas propagation kernel.\footnote{In the standard formulation, cost minimization implies that the
Cobb--Douglas exponents can be interpreted as buyer-side network weights, corresponding to the
allocation of unit expenditure across suppliers. Sanctions, however, are more naturally represented
as restrictions imposed by sellers on the supply of inputs rather than as changes in buyers'
purchasing decisions. Note that the production effects of reduced input flows are identical regardless
of which side initiates them. We exploit this equivalence by studying the supplier-side dual, in
which weights are normalized by suppliers' allocation of output across buyers. In this
representation, sanctions appear directly as changes in sellers' allocation patterns, or,
equivalently, as changes in the weights of the associated buyer-side dual. All of this is to say that our analysis pertains to an economy with general Cobb--Douglas technologies rather than only to the much more restrictive Leontief setting.} We work with the OECD World Input--Output Table, which records transactions across many sectors and countries. The nodes of our network are therefore country--sectors, and the edges represent input flows between them. Although the framework is general enough to accommodate a wide range of global supply restrictions, in this paper we focus on two benchmark cases. The first considers a restriction in which a given foreign country--sector is prevented from selling to all sectors of a target country. The second considers a restriction in which all sectors of a sanctioning country are prevented from selling to all sectors of a target country. 

\smallskip
The restrictions take the following form. We set the weights attached to sanctioned
destinations to zero and increase the weights attached to non-sanctioned destinations so as to preserve
seller-side normalization. This normalization matters. It captures the idea that a sanction is not
the same thing as a decline in productive capacity. The sanctioning country does not produce less;
it simply redirects its output elsewhere. We then study how this sanction shock propagates through
the supply chain and how it affects economic activity in the target country.

\smallskip
Our empirical analysis focuses on the vulnerability of the Indian economy, although none of the
theoretical apparatus developed in the paper is specific to her. In the country-sector
counterfactuals, where a foreign country ceases exporting the inputs of one particular sector to
India, the Indian economy is most vulnerable to the oil and gas sectors of Saudi Arabia and the
United Arab Emirates, followed by Australia's coal mining and China's chemicals.
Indeed, nearly all of the top twenty entries, out of more than 3,500 country-sectors, belong to
fossil fuels, mining, and related upstream activities, with Singapore's water transport sector being the primary exception. No European country-sector appears in the top twenty. This pattern is broadly preserved when we turn to country-wide sanctions, in which all sectors of a
foreign country cease supplying intermediate inputs to India. Saudi Arabia, the United Arab Emirates, and Russia figure in the top five, with China leading the group by far and the United States ranking third.  The main difference between country-sector vulnerabilities and country vulnerabilities is that now some advanced European economies enter the ranking, including the United Kingdom, Germany, and France; Japan too makes it to the list.

\subsection{Related Literature}
Economists have long grappled with the propagation and impact of shocks that originate outside the
economic system, and the first formal rendering of the idea was given by Ragnar Frisch in his
aptly titled article ``Propagation Problems and Impulse Problems in Dynamic Economics.'' Frisch
argued that the question of the origin of the impulse can be treated independently of the question
of how it propagates through the economic system. This Frischian dichotomy has since become a staple
of macroeconomic theorizing. The recent literature on the propagation of shocks through production
networks, to which our paper is closely connected, is no exception. Much of that literature is
concerned with how shocks---productivity changes, natural disasters, and others of that
kind---move through a fixed kernel. The conceptual problem posed by
sanctions is however of a wholly different kind. Here, the shock is a change in the kernel itself, more
specifically, a change in the weights of the links in the production network. We therefore have had
to develop somewhat novel ways of studying a kernel change as a shock. The closest antecedents to our approach to
kernel change are perhaps the regional extraction method and the field-of-influence approach in
input--output analysis \citep{dietzenbacher1993,caslerhadlock1997,sonishewings1992,sonishewings1998},
which study how removing a sector or region, or perturbing selected coefficients, changes the
Leontief inverse and the implied structure of interdependence.

\smallskip
Naturally, our paper fits within the literature on geoeconomics, particularly the strand that
examines how trade relations can become a source of political vulnerability.\footnote{See, for
instance, \citet{baldwin1985,hufbauerschottelliottoegg2007,drezner1999,afesorgbormahadevan2019,
felbermayr2020,kwonsyropoulosyotov2024,ghironi2025}, among others.} This literature argues that
international commerce is not merely a source of gains from trade. It can also generate asymmetries
of dependence that may be used as instruments of coercion \citep{hirschman1945}. That line of
thinking naturally leads to measuring vulnerability in terms of bilateral trade shares, import
concentration, and other indicators of aggregate exposure. These are no doubt useful measures, but
they remain largely measures of dependence. In a network setting,
however, dependence need not coincide with risk. Our paper therefore furthers the
geoeconomics literature by moving from measures of exposure to measures of risk using empirically
calibrated counterfactual experiments on the world supply chain.

\subsection{Organization of the paper}

The paper is organized as follows. Section \ref{sec:model_supply_disruptions} presents the basic model of the world supply chain and describes how disturbances dissipate through the system. Section \ref{sec:supply_restrictions} explains how sanctions are imposed by reforming the weights of the world production network. It also characterizes the propagation of a shock to the world supply-chain kernel. Section \ref{sec:empirical} calibrates the model to the OECD World Input--Output Table and runs counterfactual experiments on sanctions by foreign country-sectors in order to estimate the vulnerability of the Indian economy to supply restrictions. Section \ref{sec:conclusion} offers concluding remarks.

\section{A Model of Supply Disruptions}
\label{sec:model_supply_disruptions}

\noindent
\emph{Notation.} Throughout, matrices are denoted by uppercase bold letters, vectors by lowercase
bold letters, and scalars by lowercase letters: \(\mathbf{X}\) denotes a matrix, \(\mathbf{x}\) a
vector, and \(x\) a scalar. We use ordinary superscripts for powers, as in \(x^t\), while
subscripts, as in \(x_t\), and parenthesized superscripts, as in \(x^{(t)}\), are reserved for
indexing.

\subsection{The network}

We model the world economy as a network of country--sector pairs. There are \(C\) countries,
indexed by \(c=1,\dots,C\), and each country contains \(S\) sectors, indexed by \(s=1,\dots,S\).
The total number of country--sector pairs is therefore
\[
N=CS
\]

For simplicity, we collapse the pair \((c,s)\) into a single index
\[
\kappa(c,s):=(c-1)S+s \in \{1,\dots,N\},
\qquad
c=1,\dots,C,\quad s=1,\dots,S
\]
Naturally, the first \(S\) entries correspond to the sectors of country \(1\), the next \(S\)
entries to the sectors of country \(2\), and so on, with
\[
\kappa(1,1)=1
\qquad\text{and}\qquad
\kappa(C,S)=N
\]

We denote countries \(i\) and \(j\) by \(c_i\) and \(c_j\), and country--sectors \(i\) and \(j\)
by \(\kappa_i\) and \(\kappa_j\). We order the country--sector pairs first by country, and then by
sector within country. This induces a block structure on the adjacency matrix of the world
input--output network. More specifically, the adjacency matrix \(\mathbf A\) has a
\(C\times C\) block structure, with each block of dimension \(S\times S\),
\begin{equation}
\label{eq:matrix}
\mathbf A=
\begin{bmatrix}
\mathbf A_{11} & \mathbf A_{12} & \cdots & \mathbf A_{1C}\\
\mathbf A_{21} & \mathbf A_{22} & \cdots & \mathbf A_{2C}\\
\vdots & \vdots & \ddots & \vdots\\
\mathbf A_{C1} & \mathbf A_{C2} & \cdots & \mathbf A_{CC}
\end{bmatrix},
\qquad
\mathbf A_{c_i c_j}\in\mathbb R_+^{S\times S}
\end{equation}

Each block \(\mathbf A_{c_i c_j}\) records sector-to-sector allocation shares from supplier country
\(c_j\) to user country \(c_i\). The diagonal blocks \(\mathbf A_{c_i c_i}\) therefore capture
within-country production linkages, while the off-diagonal blocks capture cross-country input
dependence. Let \(\mathbf A=[a_{ij}]_{i,j=1}^N\) denote the matrix of intermediate-input allocation,
where \(a_{ij}\) is the share of the output of country-sector \(\kappa_j\) allocated to
country-sector \(\kappa_i\). We normalize these allocation shares so that
\begin{equation}
\sum_{i=1}^N a_{ij}=1
\qquad\text{for all } j=1,\dots,N
\label{eq:A_colsum_one}
\end{equation}
Thus, each column of \(\mathbf A\) describes how the output of a given seller country-sector is
distributed across destination country-sectors.

\smallskip
Not all output produced by a country-sector remains within the intermediate-input network. A
sizeable share leaks out for final consumption, inventories, wastage, and other uses outside the
production loop, and these leakage rates vary across country-sectors. While \(\mathbf A\) captures
how the within-network component of output is allocated across downstream country-sectors, it does
not by itself account for the fact that only part of total output is available for use as an
intermediate input. To incorporate these leakages into our analysis, let \(\beta_j\in(0,1)\)
denote the leakage rate of country-sector \(\kappa_j\), that is, the share of its output that exits
the intermediate-input network. We collect these into the diagonal matrix
\begin{equation}
\boldsymbol\beta:=\operatorname{diag}(\beta_1,\dots,\beta_N).
\label{eq:beta_diag_def}
\end{equation}
Then \(\mathbf I-\boldsymbol\beta\) is the diagonal matrix of within-network retention rates, whose
\(j\)-th diagonal entry is \(1-\beta_j\)

We therefore define the benchmark propagation kernel by
\begin{equation}
\mathbf K:=\mathbf A(\mathbf I-\boldsymbol\beta)
\label{eq:K_def_revised}
\end{equation}
Its \((i,j)\)-th entry is
\[
K_{ij}=a_{ij}(1-\beta_j)
\]
Thus \(\mathbf A\) determines where supplier \(j\)'s within-network output is directed, while
\(1-\beta_j\) determines how much of its total output remains available for propagation through the
network in the first place. Column \(j\) of \(\mathbf K\) therefore gives the effective
distribution of goods from supplier \(j\) across downstream users after accounting for leakage.

\subsection{Existence and stability of equilibrium}

We assume throughout that the benchmark propagation kernel \(\mathbf K\) is irreducible and
aperiodic. Thus the world production network forms a single interconnected system, and its
propagation dynamics are not confined to disconnected components or purely cyclical classes. Under these assumptions, the benchmark network admits a well-defined and stable
equilibrium. The proof of which is rather simple. Note that the sum of column \(j\) of \(\mathbf K\) is
\[
\sum_{i=1}^N K_{ij}
=
\sum_{i=1}^N a_{ij}(1-\beta_j)
=
(1-\beta_j)\sum_{i=1}^N a_{ij}
=
1-\beta_j
<1
\]
Hence \(\mathbf K\) is strictly substochastic. In particular,
\[
\|\mathbf K\|_1
=
\max_j \sum_{i=1}^N K_{ij}
=
\max_j(1-\beta_j)
<1
\]
and therefore
\[
\rho(\mathbf K)\le \|\mathbf K\|_1<1
\]
where \(\rho(\mathbf K)\) denotes the spectral radius of \(\mathbf K\).

\smallskip
Since \(\rho(\mathbf K)<1\), the matrix \(\mathbf I-\mathbf K\) is invertible, so the benchmark
network has a unique finite equilibrium. The same condition also implies stability:
\[
\mathbf K^t\to \mathbf 0
\qquad\text{as } t\to\infty
\]
Hence any disturbance propagated through the network eventually dies out. The equilibrium is
therefore globally stable, and \(\rho(\mathbf K)\) governs the asymptotic rate at which shocks
dissipate.

\section{Supply Restrictions}
\label{sec:supply_restrictions}

We model a foreign supply restriction as a perturbation to the benchmark allocation matrix
\(\mathbf A\in\mathbb R_+^{N\times N}\), holding fixed the leakage rates collected in
\(\boldsymbol\beta\). Throughout, the \emph{source country} denotes the country imposing the
restriction, and the \emph{target country} denotes the country on which the restriction is imposed.
The source country may restrict the flow of inputs from either all of its sectors or a selected
subset of its sectors to the target country.\footnote{In the baseline setting, we take the
restriction to apply to all sectors of the target country, which captures many sanctions regimes.
The framework can be adapted readily to narrower restrictions that apply only to particular sectors
or technologies.} Since the source country is not assumed to lose productive capacity, the
curtailed sales are not destroyed; instead, the unsold output is redirected to other buyers. A
restriction therefore changes the allocation weights of the relevant supplier country-sectors:
weights on the restricted target are set to zero, while weights on unrestricted destinations are
increased so as to preserve seller-side normalization.

We now formalize this disruption and renormalization. Fix a source country and a target country.
Since rows of \(\mathbf A\) index users and columns index suppliers, a restriction aimed at the
target country acts on the rows corresponding to that country's sectors. For each country-sector
\(i=1,\dots,N\), define
\[
q_i=
\begin{cases}
1, & \text{if country-sector }i\text{ belongs to the target country}\\
0, & \text{otherwise}
\end{cases}
\]
and
\[
r_i=
\begin{cases}
1, & \text{if country-sector }i\text{ is a restricted sector in the source country}\\
0, & \text{otherwise}
\end{cases}
\]
Thus \(q_i\) indicates whether row \(i\) belongs to the target country, while \(r_i\) indicates
whether supplier column \(i\) is subject to the restriction. We then collect these indicators into
the diagonal matrices
\[
\mathbf Q:=\operatorname{diag}(q_1,\dots,q_N)\in\mathbb R^{N\times N},
\qquad
\mathbf R:=\operatorname{diag}(r_1,\dots,r_N)\in\mathbb R^{N\times N}
\]
Thus \(\mathbf Q\) selects the target-country rows, while \(\mathbf R\) selects the supplier
sectors in the source country whose sales to the target country are restricted. Depending on the
application, \(\mathbf R\) may select a single sector of the source country, several sectors, or
all sectors of the source country. We do not introduce separate superscript notation for these
cases; the intended case will be clear from the choice of \(\mathbf R\) and the surrounding
discussion.

For notational ease, whenever a matrix carries the superscript \(^{\circ}\), it denotes the
post-restriction object associated with a particular choice of source country, target country, and
selector matrix. We suppress this dependence unless it is needed for clarity.

For a given selector \(\mathbf R\), define the pre-renormalization restricted matrix
\begin{equation}
\boldsymbol{\mathcal A}
:=
\mathbf A-\mathbf Q\mathbf A\mathbf R,
\qquad
\boldsymbol{\mathcal A}\in\mathbb R^{N\times N}
\label{eq:B_general_supply}
\end{equation}
Thus, for each supplier column selected by \(\mathbf R\), the entries corresponding to the target
country are set to zero, while all other entries remain unchanged.

For each supplier column \(\ell\), define
\begin{equation}
\psi_\ell
:=
\sum_{i=1}^N q_i a_{i\ell},
\qquad
\psi_\ell\in[0,1]
\label{eq:psi_general_supply}
\end{equation}
Thus \(\psi_\ell\) is the benchmark share of supplier \(\ell\)'s within-network deliveries directed
to the target country. Then, if \(r_\ell=1\), column \(\ell\) of \(\boldsymbol{\mathcal A}\) sums
to \(1-\psi_\ell\); if \(r_\ell=0\), it continues to sum to \(1\). Thus only the affected supplier
columns lose mass, necessitating renormalization.

We implement this counterfactual as reallocative rather than destructive. Output that can no longer
be sent to the target country is redirected across unrestricted destinations in proportion to their
benchmark weights. To impose this normalization, define
\begin{equation}
\mathbf N
:=
\operatorname{diag}(n_1,\dots,n_N)\in\mathbb R^{N\times N}
\label{eq:N_general_supply}
\end{equation}
where
\begin{equation}
n_\ell=
\begin{cases}
\displaystyle \frac{1}{1-\psi_\ell}, & r_\ell=1\\[0.8em]
1, & r_\ell=0
\end{cases}
\label{eq:n_general_supply}
\end{equation}
This requires \(1-\psi_\ell>0\) for every restricted column \(\ell\). The renormalized restricted
allocation matrix is then
\begin{equation}
\mathbf A^{\circ}
:=
\boldsymbol{\mathcal A}\mathbf N,
\qquad
\mathbf A^{\circ}\in\mathbb R^{N\times N}
\label{eq:Astar_general_supply}
\end{equation}
Right multiplication by \(\mathbf N\) rescales each affected supplier column separately, restoring
its column sum to one while preserving the relative weights across unrestricted destinations. We
refer to \(\mathbf A^{\circ}\) as the restricted allocation matrix.

Equivalently, if \(r_\ell=1\), then the entries of the restricted matrix satisfy
\[
(\mathbf A^{\circ})_{i\ell}
=
\begin{cases}
0,
& \text{if country-sector }i\text{ belongs to the target country}\\[0.8em]
\displaystyle \frac{a_{i\ell}}{1-\psi_\ell},
& \text{otherwise}
\end{cases}
\]
while all unselected columns remain unchanged.

Since the leakage structure is held fixed, the counterfactual propagation kernel associated with the
restriction is
\begin{equation}
\mathbf K^{\circ}
:=
\mathbf A^{\circ}(\mathbf I-\boldsymbol\beta)
\label{eq:Kstar_general_supply}
\end{equation}

\subsection{Propagation of a sanction shock}
\label{subsec:propagation_supply_restrictions}

The economic effect of a supply restriction is not exhausted by the immediate loss of a set of
bilateral input flows. Because those flows are embedded in a production network, the restriction
alters the subsequent rounds by which disturbances are transmitted through the economy. The relevant
object is therefore not merely the direct disappearance of some links, but the change in the
propagation structure itself and the cumulative consequences of that change over time. We now formalize these dynamics. 

\smallskip
Note that the primitive
sanction shock is thus the kernel perturbation
\begin{equation}
\Delta\mathbf K
:=
\mathbf K^{\circ}-\mathbf K
\label{eq:DeltaK_general_supply}
\end{equation}
This matrix captures the direct change in the one-step transmission structure induced by the
restriction.

To understand the dynamic consequences of this shock, note that under the benchmark network, \(t\)
rounds of propagation are governed by \(\mathbf K^t\), whereas under the restricted network they are
governed by \((\mathbf K^{\circ})^t\). The restriction therefore changes not only the first round
of transmission, but the entire propagation sequence. For a finite horizon \(T\geq 1\), define the
corresponding change in cumulative propagation by

\begin{align}
\boldsymbol\Phi_{\mathrm T}
&:=
\sum_{t=0}^{T-1}\mathbf K^t - \sum_{t=0}^{T-1}(\mathbf K^{\circ})^{t}
\label{eq:Phi_T_general_supply}
\end{align}

Its \((k,\ell)\)-th entry records the change in cumulative transmission from origin country-sector
\(\ell\) to destination country-sector \(k\) up to horizon \(T\). Our main object of interest,
however, is the long-run counterpart obtained as \(T\to\infty\). Under the stability conditions
imposed earlier,
\begin{equation}
\boldsymbol\Phi
:=
\lim_{T\to\infty}\boldsymbol\Phi_{\mathrm T}
=
(\mathbf I-\mathbf K)^{-1} - (\mathbf I-\mathbf K^{\circ})^{-1}
\label{eq:Phi_general_supply}
\end{equation}
This is the difference between the long-run propagation operators of the restricted and benchmark
economies. It summarizes how the restriction changes total cumulative transmission once all rounds
of propagation are taken into account.

\section{Empirical Assessment of the Impact of Foreign Supply Restrictions on the Indian Economy}
\label{sec:empirical}

\subsection{Statistics on the impact of supply restrictions}
\label{subsec:country_level_statics}

We now translate the long-run propagation effect \(\boldsymbol\Phi\) into country-level statistics.
Since the source country and the target country are fixed throughout, we simplify notation
accordingly. The object \(\boldsymbol\Phi\) is defined at the country-sector level: its rows and
columns are indexed by individual country-sectors. Our aim in this section is to aggregate the
effect on the target side across all sectors of the target country, and then across the relevant
source-country sectors to which the restriction applies.

Let
\[
\boldsymbol\mu=(\mu_1,\dots,\mu_N)^\top\in\{0,1\}^N
\]
denote the selector vector for the sectors of the target country, with
\[
\mu_i=
\begin{cases}
1, & \text{if country-sector }i\text{ belongs to the target country}\\
0, & \text{otherwise}
\end{cases}
\]
Thus \(\boldsymbol\mu\) has ones on the sectors of the target country and zeros elsewhere. Left
multiplication by \(\boldsymbol\mu^\top\) therefore aggregates across all sectors of the target
country. 

Let
\[
\boldsymbol\nu=(\nu_1,\dots,\nu_N)^\top\in\{0,1\}^N
\]
denote the selector vector for the source-country sectors covered by the restriction, with
\[
\nu_i=
\begin{cases}
1, & \text{if country-sector }i\text{ is a restricted sector in the source country}\\
0, & \text{otherwise}
\end{cases}
\]
Thus \(\boldsymbol\nu\) has ones on the restricted sectors of the source country and zeros
elsewhere. Right multiplication by \(\boldsymbol\nu\) therefore selects, and then sums over, the
source-country sectors through which the restriction operates. Depending on the case under
consideration, \(\boldsymbol\nu\) may select a single source-country sector, several
source-country sectors, or all sectors of the source country.

Under the benchmark network, define
\begin{equation}
\mathbf y
:=
\boldsymbol\mu^{\top}(\mathbf I-\mathbf K)^{-1}
\label{eq:y_kernel_def}
\end{equation}
and under the restricted network define
\begin{equation}
\mathbf y^{\circ}
:=
\boldsymbol\mu^{\top}(\mathbf I-\mathbf K^{\circ})^{-1}
\label{eq:y_kernel_restricted_def}
\end{equation}
These are row vectors indexed by origin country-sectors. Their \(\ell\)-th entries measure the
total long-run effect on the target country---summed across all sectors of the target country---of a
unit impulse originating in country-sector \(\ell\), under the benchmark and restricted networks,
respectively.

Their difference,
\begin{equation}
\Delta \mathbf y
:=
\mathbf y - \mathbf y^{\circ}
=
\boldsymbol\mu^{\top}\boldsymbol\Phi
\label{eq:Delta_yij}
\end{equation}
therefore records how the restriction changes the target country's long-run exposure to impulses
originating in each country-sector.

Let
\begin{equation}
\mathbf Z:=\diag(z_1,\dots,z_N)
\label{eq:Z}
\end{equation}
be the diagonal matrix of country-sector scale, where \(z_\kappa>0\) denotes the size of
country-sector \(\kappa\).

\begin{defini}[Vulnerability Index]
Define the vulnerability index associated with the restriction by
\begin{equation}
\gamma
:=
\Delta \mathbf y\,\mathbf Z\,\boldsymbol\nu
\label{eq:gamma_general}
\end{equation}
\end{defini}

\(\gamma\) measures the size-weighted change in the
target country's long-run exposure to the restricted sectors of the source country.

\subsection{Calibration of the model}

\begin{figure}[h!]
\centering
\scalebox{0.8}{\includegraphics[width=1\textwidth]{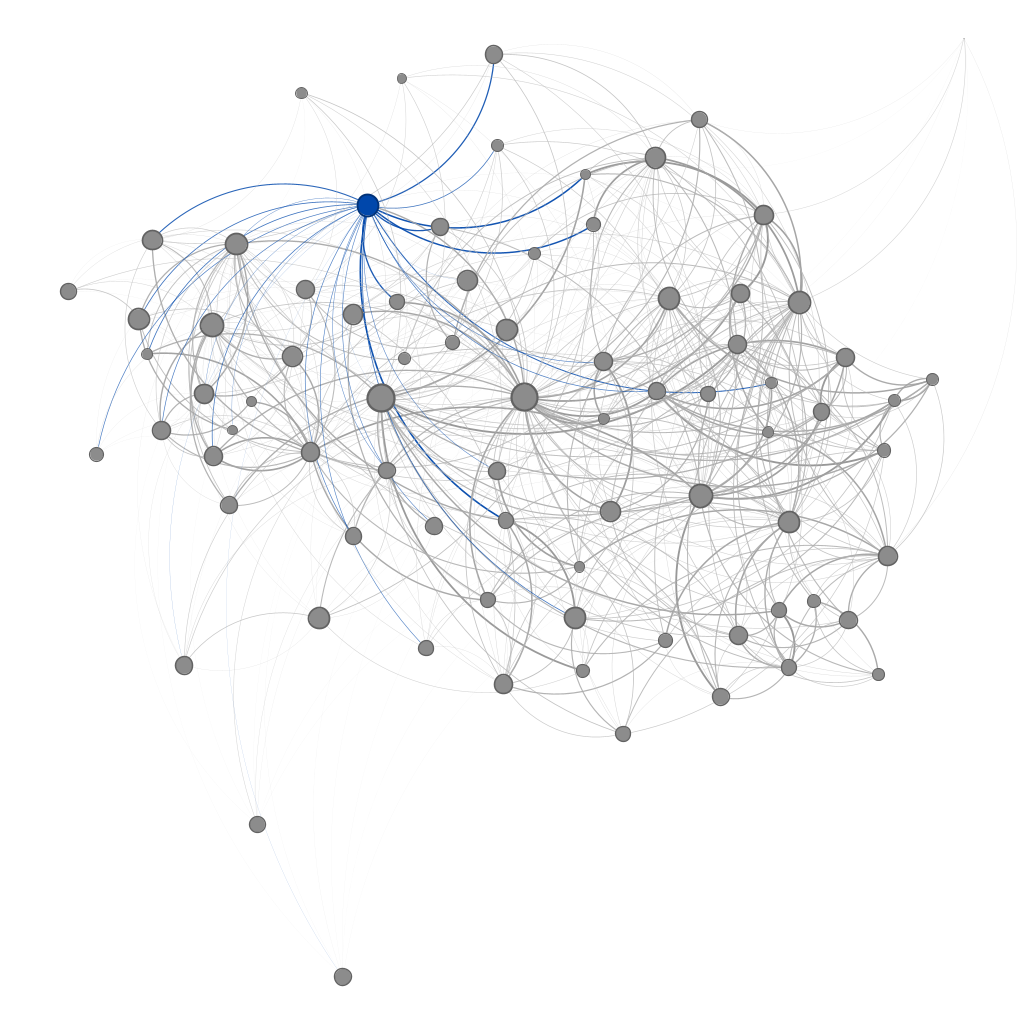}}
\caption{The network of intermediate-input flows between countries. The graph contains the flows of
80 countries in the OECD input-output table. Node sizes reflect the total production of
intermediate inputs by each country. India is marked in blue, as are the flows of intermediate
inputs into India.}
\label{fig:gephi}
\end{figure}

Our empirical analysis is based on the Inter-Country Input--Output (ICIO) tables of
\citet{oecdicio}. We use the 2005 table, which in our data covers 80 economies and 45 sectors. The
ICIO tables provide a global representation of cross-country and cross-industry transactions,
distinguishing intermediate uses from non-intermediate uses. Figure~\ref{fig:gephi} presents a
visualization of these intermediate-input flows after aggregation to the country level. In this
graph, each node represents a country, and each directed link represents the value of intermediate
inputs supplied from one country to another. Node size is proportional to the total production of
intermediate inputs by that country. The figure makes clear two features of the data that motivate
the subsequent analysis. First, the world economy is organized as a dense but highly uneven network
of intermediate-input linkages, with some countries occupying much more central positions than
others. Second, India is embedded in this network through a wide range of foreign suppliers, but the
intensity of these links varies sharply across countries. Our empirical analysis exploits precisely
this structure: it uses the observed country-sector input-output flows to quantify how disruptions
originating in particular foreign sectors, or in all sectors of a foreign country, propagate through
the production network and affect Indian economic activity.

\smallskip
We calibrate the primitive objects introduced earlier in the paper directly from these data. Let
\(x_{ij}\) denote the observed intermediate flow from supplier country-sector \(j\) to user
country-sector \(i\). We define the size of country-sector \(j\) by its total intermediate output,
\[
z_j:=\sum_{i=1}^N x_{ij}
\]
and collect these into the diagonal matrix $\mathbf Z$.

\smallskip
From the same intermediate-flow matrix we construct the empirical counterpart of the benchmark
allocation matrix \(\mathbf A=[a_{ij}]_{i,j=1}^N\) introduced in \eqref{eq:matrix}, with
\[
a_{ij}
=
\frac{x_{ij}}{z_j}
=
\frac{x_{ij}}{\sum_{r=1}^N x_{rj}}
\]
so that each column of \(\mathbf A\) sums to one whenever country-sector \(j\) has positive
intermediate output.  We next calibrate the leakage rates introduced earlier in the model. For each country-sector \(j\),
the leakage rate \(\beta_j\) is computed as the share of output flowing to non-intermediate uses,
that is, to final demand and other uses outside the intermediate-input network. Collecting these
terms into the diagonal matrix \(\boldsymbol\beta\) defined in \eqref{eq:beta_diag_def}, we obtain
the empirical counterpart of the benchmark propagation kernel $\mathbf K$ defined in
\eqref{eq:K_def_revised}.

\smallskip
Throughout this section, India is the target country. We consider two classes of counterfactual
foreign supply restrictions. The first consists of sector-specific restrictions. For each foreign
country-sector, we impose a complete cessation of intermediate-input flows from that
country-sector to all sectors in India, reconstruct the corresponding restricted propagation kernel, and compute the resulting effect on the Indian economy.  Since the database contains 79 foreign source countries and 45 sectors
per country, this yields \(79\times 45=3{,}555\) country-sector counterfactuals.

\smallskip
The second class consists of country-wide restrictions. For each foreign country, we impose a
complete cessation of intermediate-input flows from all sectors of that country to all sectors in
India, reconstruct the corresponding restricted propagation kernel, and
compute the resulting effect on the Indian economy. Since India is fixed as the destination, this
yields 79 country-level counterfactuals.

\smallskip
For each counterfactual, we compute the vulnerability index
\(\gamma\) defined in \eqref{eq:gamma_general}, where the selector vector \(\boldsymbol\nu\)
identifies the source-country sectors covered by the restriction. In the sector-specific exercises,
\(\boldsymbol\nu\) selects one foreign country-sector. In the country-wide exercises,
\(\boldsymbol\nu\) selects all sectors of the foreign source country. Thus the same statistic
\(\gamma\) applies to both classes of experiments; only the choice of \(\boldsymbol\nu\) differs
across counterfactuals.

To facilitate comparison within each class of experiment, we normalize the reported indices by
dividing by the largest observed value in that class, so that the maximum normalized score is equal
to one. A value close to one therefore identifies a foreign country or foreign country-sector whose
restriction is among the most disruptive for India within that class. This normalization preserves
the underlying ranking while making magnitudes easier to compare across tables and figures.
Section~\ref{subsec:country_sec} presents the results for sector-specific restrictions, while
Section~\ref{subsec:country} presents the corresponding results for country-wide restrictions.

\subsection{Country-sector vulnerability}
\label{subsec:country_sec}

\begin{figure}[H]
\centering
\scalebox{0.9}{\includegraphics[width=1\textwidth]{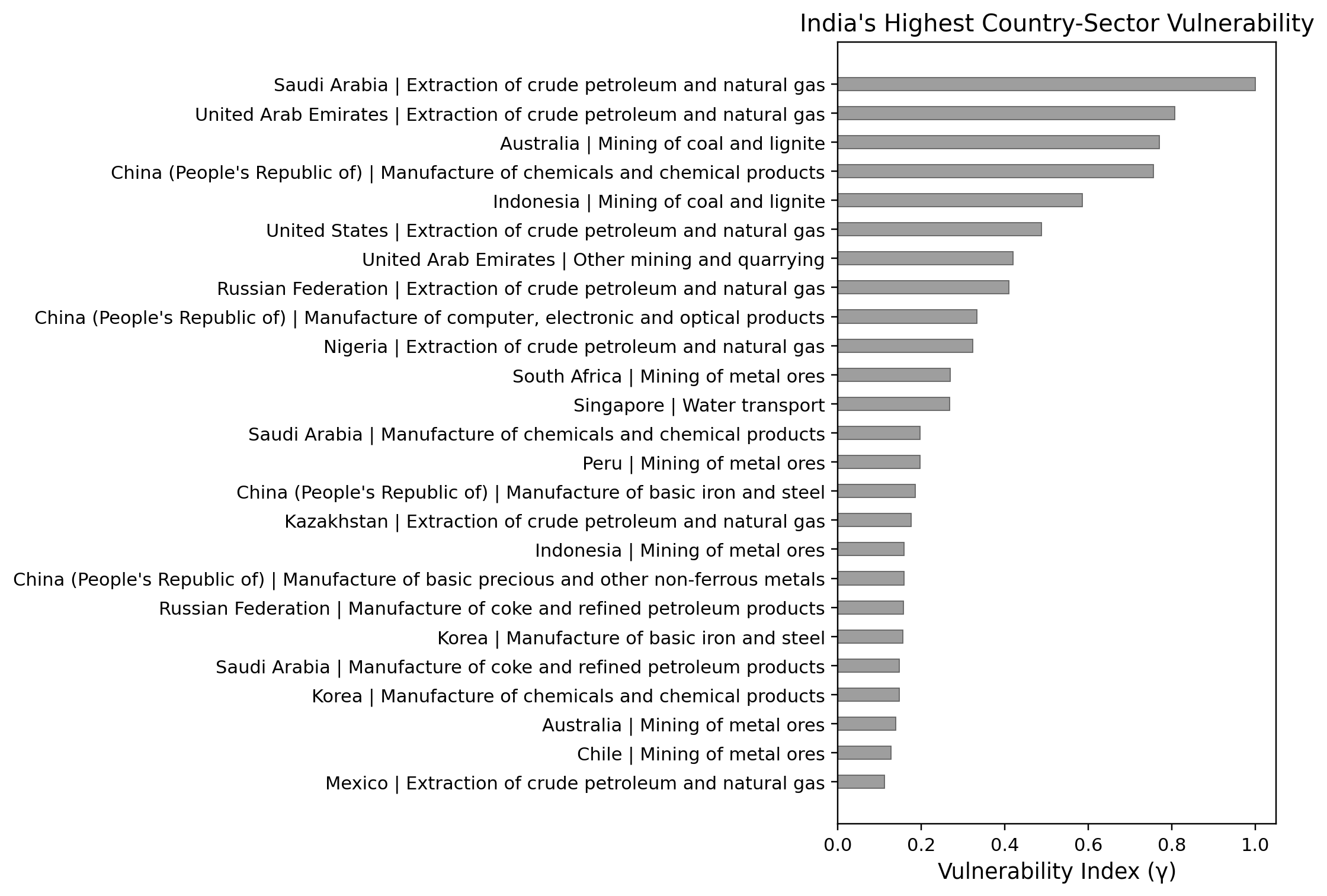}}
\caption{India's vulnerability to the closure of intermediate-input flows
from specific foreign country-sectors.}
\label{fig:sec_hist}
\end{figure}

\begin{figure}[H]
\centering
\scalebox{0.9}{\includegraphics[width=1\textwidth]{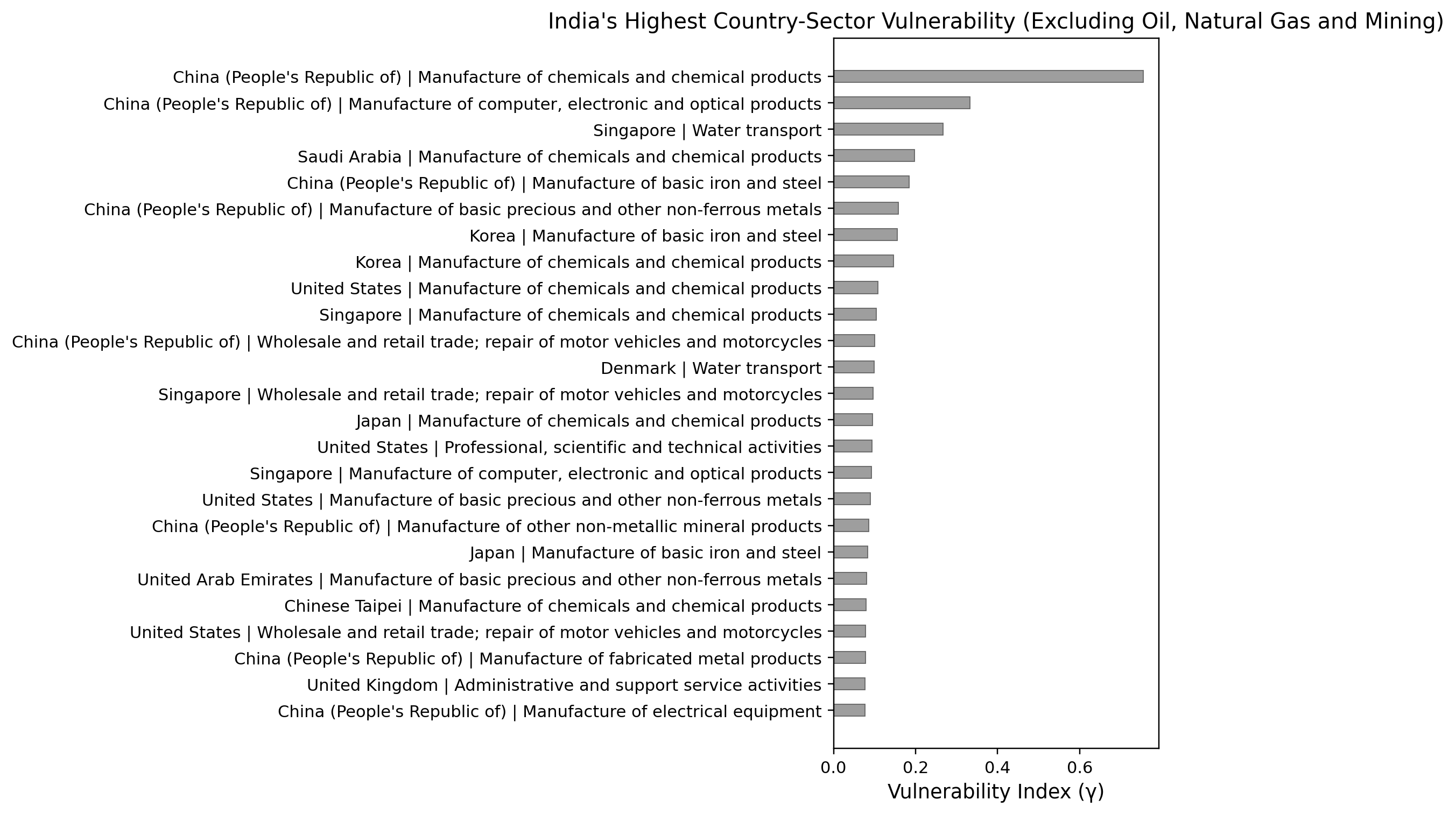}}
\caption{India's vulnerability to the closure of intermediate-input flows
from specific foreign country-sectors, excluding sectors that produce oil, gas, and mining.}
\label{fig:sec_hist_short}
\end{figure}

Figure~\ref{fig:sec_hist} ranks the foreign country-sectors that generate the greatest
vulnerability for the Indian economy. The two largest vulnerabilities arise from the extraction of
petroleum and natural gas in Saudi Arabia and the United Arab Emirates, followed closely by coal
mining in Australia and chemicals in China. Figure~\ref{fig:sec_hist_short} reports the same
ranking after excluding oil \& gas production and mining. The broad pattern remains unchanged: India's economy
appears especially vulnerable to foreign supplies of chemicals and metals,
rather than to high-end technology sectors. There are exceptions: the most prominent being `professional, scientific, and technical activities' of the United States.  It is striking that sectors from the most European economies are largely absent, even after removing oil, gas, and mining. 

\begin{figure}[H]
\centering
\scalebox{0.9}{\includegraphics[width=1\textwidth]{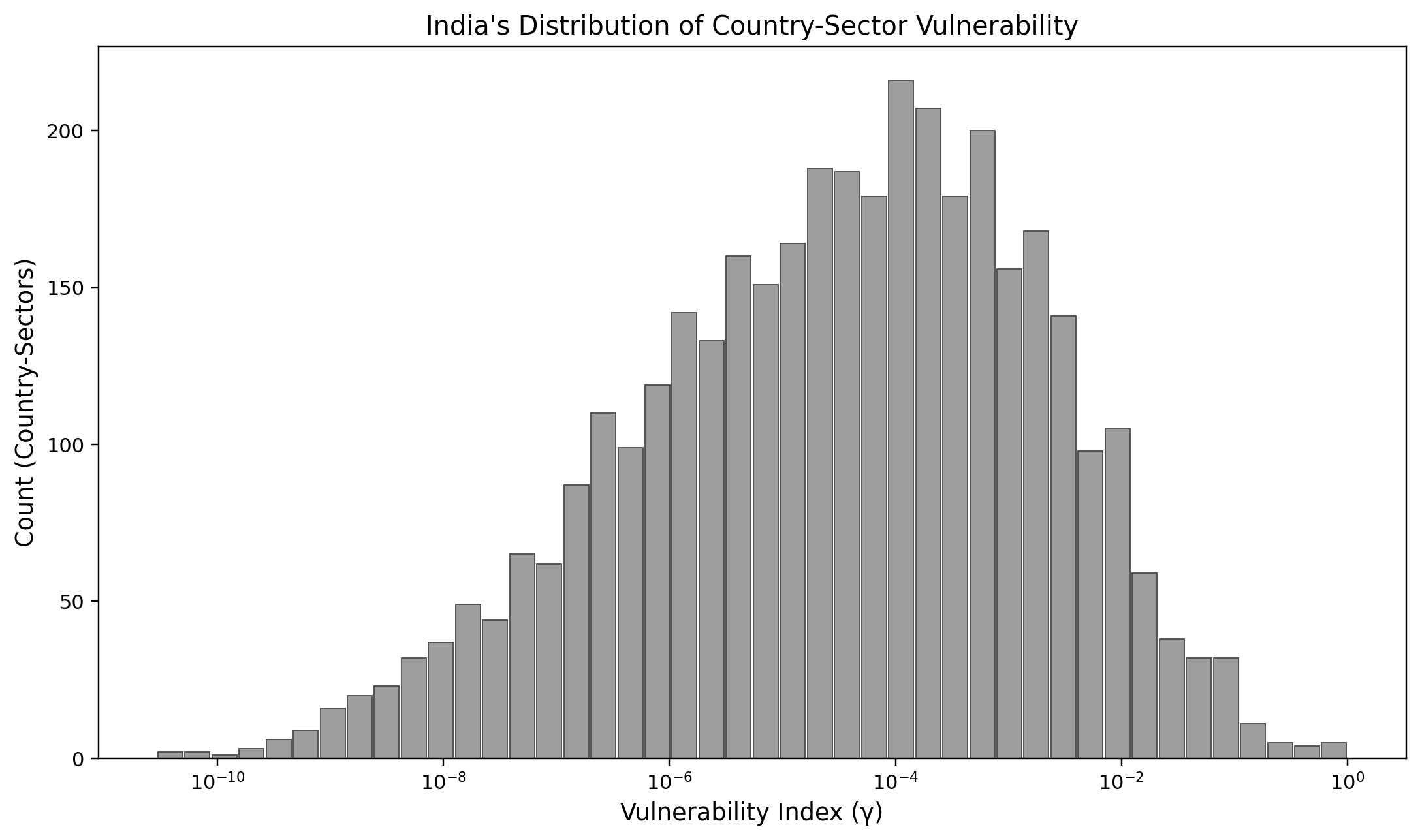}}
\caption{The distribution of India's vulnerability to the complete closure of
intermediate-input flows from specific foreign country-sectors.}
\label{fig:sec_dist}
\end{figure}

\smallskip
Figure~\ref{fig:sec_dist} displays the distribution of India's vulnerability across foreign
country-sectors. Note that the vulnerability index plotted on a logarithmic horizontal axis, which makes  makes the extreme skewness of the distribution apparent: a very small number of foreign
country-sectors account for a disproportionate share of India's vulnerability, while the vast
majority contribute very little. Indeed, the top 10 country-sectors account for about a third of total
vulnerability, and the top 100 for nearly three-quarters. This in a universe of more than 3{,}500 country-sectors.

\smallskip
The main empirical message is therefore one of strong concentration. India's exposure to foreign
supply disruptions is not broadly dispersed across the world economy, but instead heavily clustered
in a narrow set of upstream sectors tied to energy, mining, and basic industrial inputs. In this
sense, the Indian economy is vulnerable not only because it depends on foreign supply, but also
because that dependence is itself highly concentrated.

\subsection{Country vulnerability}
\label{subsec:country}

\begin{figure}[H]
\centering
\scalebox{1}{\includegraphics[width=1\textwidth]{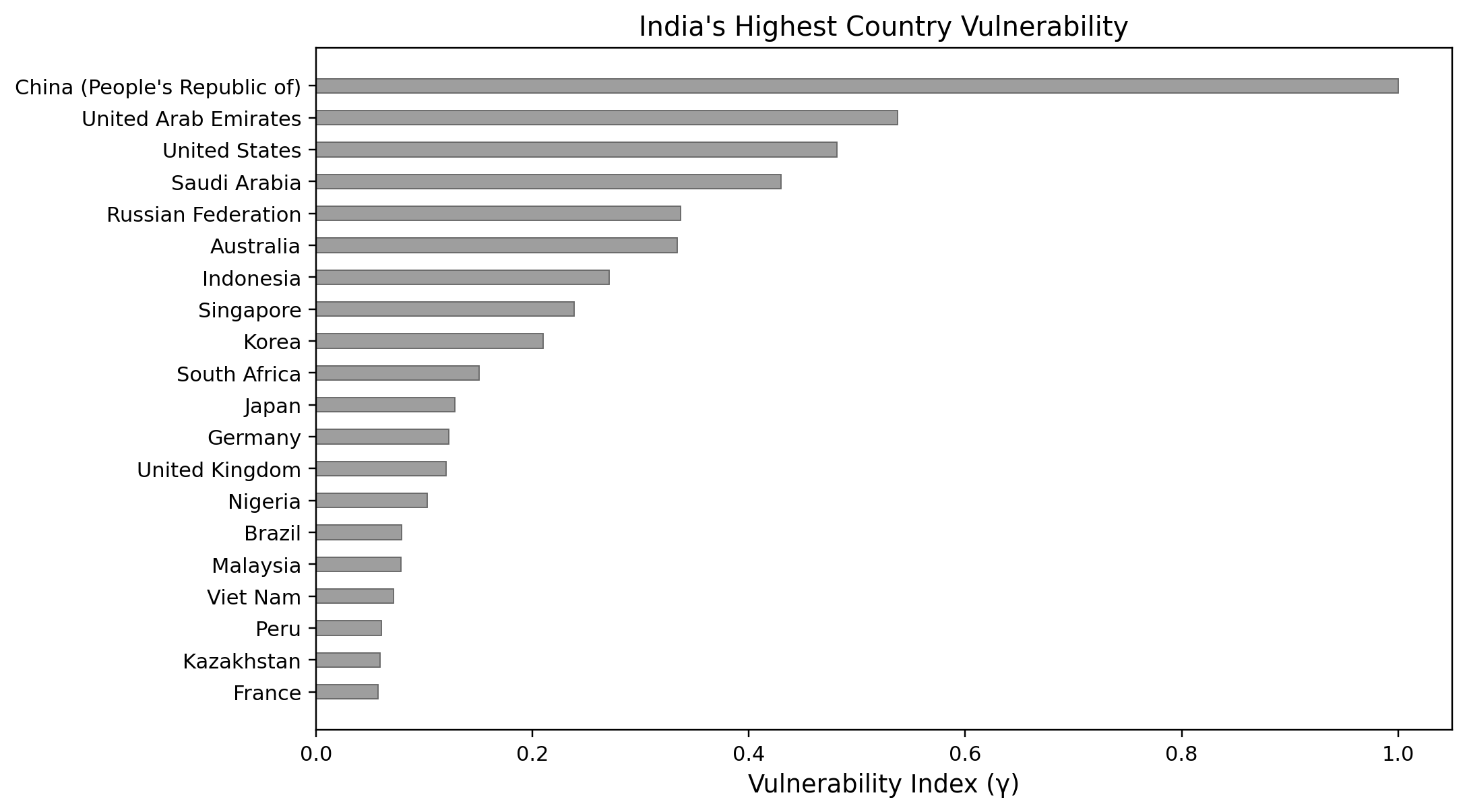}}
\caption{India's vulnerability to the complete closure of intermediate-input flows
from all sectors of a foreign country.}
\label{fig:country_sec_hist}
\end{figure}

Figure~\ref{fig:country_sec_hist} presents India's vulnerability at the country level, listing 
the twenty foreign countries whose complete interruption of intermediate-input flows would be most
disruptive for the Indian economy. Unsurprisingly, China leads the ranking, followed by the United Arab Emirates, the United States, Saudi Arabia, and Russia. This is somewhat consistent with the country-sector results in
Section~\ref{subsec:country_sec}, where India's dependence on petroleum and natural gas from these economies (except China) emerged clearly.  Unlike the country-sector ranking, however, the country-level ranking also brings advanced European economies such as Germany, the United Kingdom and France into the fray. Japan almost makes it to the top ten. 

\begin{figure}[H]
\centering
\scalebox{1}{\includegraphics[width=1\textwidth]{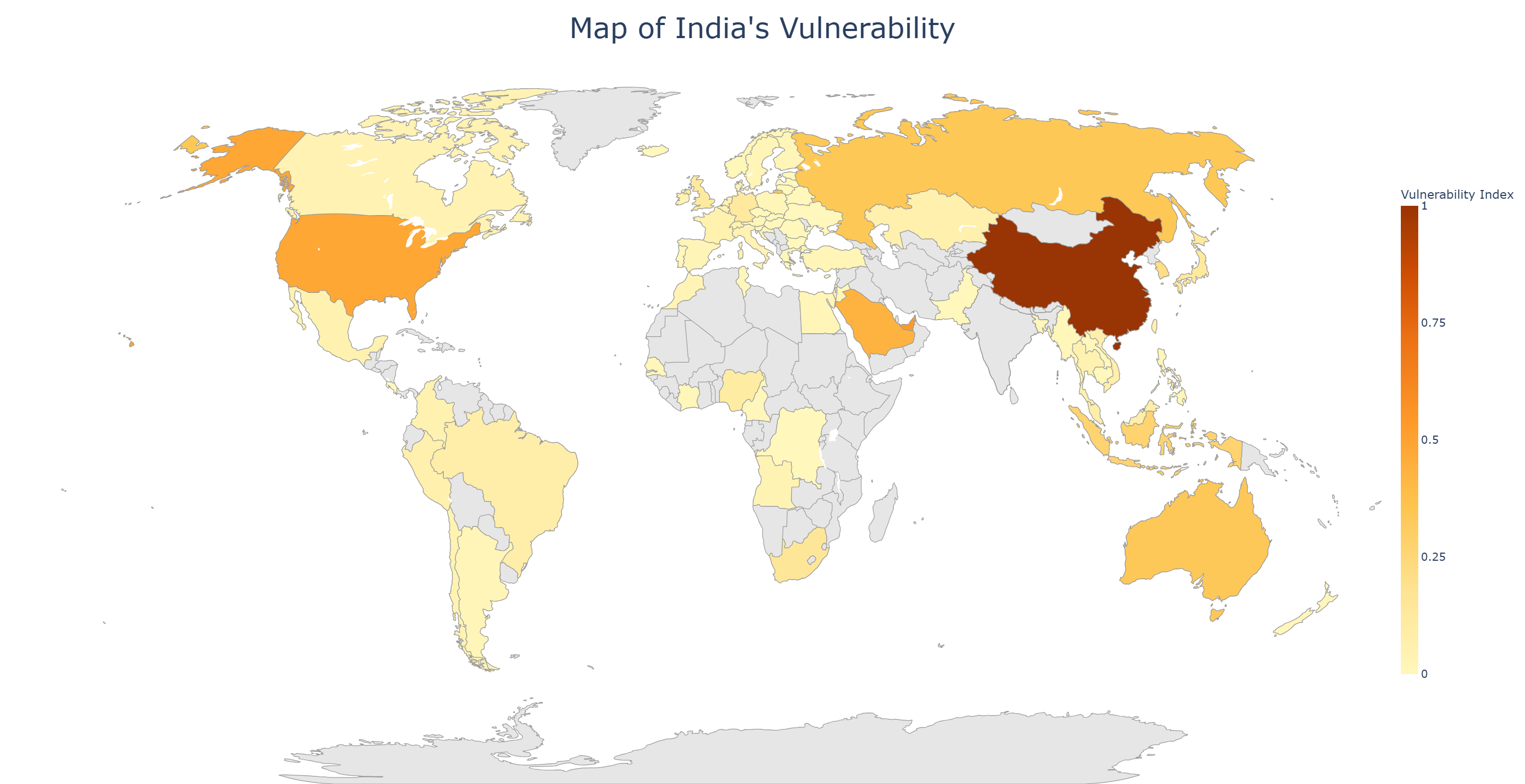}}
\caption{Vulnerability of the Indian economy under country-wide supply restrictions.}
\label{fig:map}
\end{figure}

\smallskip
Figure~\ref{fig:map} complements this ranking with a world map in which countries are shaded by the
vulnerability they pose to the Indian economy under a country-wide restriction, that is, under the
complete cessation of intermediate-input flows from all of their sectors to India. The geographic
pattern confirms that India's external vulnerability is concentrated in a relatively small set of
countries, with especially strong exposure to energy and ore suppliers (with China being the exception). In fact, the top ten countries account for nearly 70\% of total vulnerability. More
generally, India's exposure is far from uniform across trading partners: some countries matter very
little in aggregate, whereas others are capable of generating substantial disruption if they curtail
the flow of intermediate inputs.

\section{Concluding Thoughts}
\label{sec:conclusion}

It is not uncommon for an undergraduate student to be introduced to the notion of `gains from trade' by way of the example of two individuals\footnote{Often with reassuringly male and female names, so that the poor eggling is not burdened with the enormous difficulty of telling the two individuals apart at every step of the analysis.}, each endowed with some assortment of fruits that they propose to exchange with one another, with the gains from trade being the extent to which their initial endowments differ from their likings for the fruits. The question of production is then pursued along much the same riverbed. The only difference is that each `economic entity' is now equipped with a technology by which one good may be transformed into another. Gains from trade, now, arise not only because a person's endowment differs from her preferences, but also because her productive powers differ from those of others.\footnote{Ricardo, of course, presents the question of international trade in something very like this way, with England and Portugal standing in for the classroom traders of fruit. The famous cloth-and-wine example in Chapter 7 of \textit{On the Principles of Political Economy and Taxation} is more dignified than apples and oranges, but the structure is much the same: two parties, two goods, and gains from specialization arising from given differences in productive possibilities. What is largely absent, however, is the idea that trade itself may widen the division of labour and thereby reshape those productive possibilities over time.}

\smallskip
The real trouble with this analytical apparatus is not that it is wrong, but that it has no proper place for what is arguably the true fountainhead of the gains from trade: not exogenously given differences in preferences or productive possibilities, but the economies of scale that endogenously generate and enlarge such differences. One sees an inkling of this in Adam Smith's remark that the division of labour is limited by the extent of the market. But perhaps the most explicit recognition of the point is found in Allyn Young's deceptively simple essay, ``Increasing Returns and Economic Progress.''\footnote{Young's point is that increasing returns do not sit quietly inside the firm like a well-behaved production coefficient. They spill outward through the division of labour, the enlargement of markets, and the cumulative reorganizations by which one improvement breeds another. Economic progress is therefore not simply a matter of doing more of the same, but of transforming the very structure within which production takes place. See Young's classic essay, \textit{Increasing Returns and Economic Progress}. For later reflections, see \citet{buchanan1994return} on `The Return of Increasing Returns'.} What Young had in mind was not merely increasing returns inside the little fenced garden of the firm, but increasing returns that spill well beyond it, into the larger economic organism of which the firm is only one small part. In such a setting, size matters. The greater the number of individuals who cooperate through the market by way of specialization and exchange, the greater the gains from trade. The trouble is that while size matters for the gains from trade, size also changes the organization of human life. The problem of organizing society so as to sustain order is very different for a society of millions of human beings than for the proverbial Crusoe and Friday bartering coconuts on a beach. With size emerges a whole array of instruments through which human conduct is shaped and curtailed, including culture, hierarchy, and coercive institutions of various kinds, among them the modern nation-state. The great gains from trade in modern market economies are therefore embedded in political structures that arise alongside the enlargement of markets. Which is to say that the dynamics of trade are almost inevitably entangled with political dynamics---unless, of course, by trade we mean nothing more than the exchange of apples for oranges in an Edgeworth Box \citep{Wagner2014Entangled,Wagner2016Politics}.

\smallskip
Individuals raised within political communities learn to salute the flutter of certain patterns, revere this tune, honor that symbol\footnote{For a sufficiently old individual, the musculoskeletal system engages towards these reveries nearly without effort.}. The `individual', as it were, is sculpted within the nation-state, and thus comes to see the world through this socially constructed window.\footnote{On the social formation of individuality, see \citet{elias1991society}, \textit{The Society of Individuals}.} The modern nation-state, no less than other enduring forms of social organization, governs through both the barrel and the psyche. Collective interests are therefore neither mystical nor disembodied; they are the socially produced sense of belonging through which individuals come to identify with one another.  Such collective interests have a law of motion of their own. They are not bound by commercial interests.  Human socities of the scale we inhabit generate immensely powerful commercial forces through the increasing returns made possible by specialization among billions, but they also generate intense political forces, with no guarantee of harmony between the two.  This means a country may impose sanctions on another, even if the imposition were to damage the production of material goods within its own borders.  Politically imposed disruptions of trade are therefore not external disturbances to the `world economy', but one of its integral possibilities, for after all the global division of labour would be unthinkable without the order generated by the political structures within which we live and die.

\medskip
We have therefore presented a rudimentary analysis of a problem that cannot be wished away, at least not for as long as human society relies on the organizing capacity of political forces. 
While the mathematical framework we develop is quite general, the empirical analysis is only as good
as the data on which it rests, which in our case are the OECD input-output tables. It is no
simple matter to construct reliable sectoral flow data across countries. Many countries do not
regularly publish high-quality information on the sectoral composition of output, still less on the
pattern of flows across sectors and borders. In such cases, the OECD must rely, at least in part,
on structural assumptions and indirect methods that use what is known about some economies to infer
what is less well measured in others. This is not a defect peculiar to our exercise, but a
limitation that attaches more generally to any attempt to study the global production network. Our empirical results should therefore be read in that spirit: not as
precise point estimates but as rough approximations.

\newpage
\singlespacing
\footnotesize
\bibliography{ref}
\bibliographystyle{aea}

\end{document}